\documentclass[aps,prb,amssymb,showpacs,superscriptaddress,floatfix,10pt,twocolumn]{revtex4-1}

\usepackage{graphicx}
\usepackage{bm}

\usepackage{ulem}
\usepackage[usenames]{color}

\begin{document}

\title{Anisotropy-based mechanism for zigzag striped patterns in magnetic thin films}

\author{O. V. Billoni}
\affiliation{Facultad de Matem\'atica, Astronom\'{\i}a y F\'{\i}sica, Universidad Nacional de C\'ordoba, Instituto de F\'{\i}sica Enrique Gaviola (IFEG-CONICET), Ciudad Universitaria, 5000 C\'ordoba, Argentina}
\author{S. Bustingorry}
\affiliation{CONICET, Centro At{\'{o}}mico Bariloche, 8400 San Carlos de Bariloche, R\'{\i}o Negro, Argentina}
\author{M. Barturen}
\affiliation{CONICET, Centro At{\'{o}}mico Bariloche, 8400 San Carlos de Bariloche, R\'{\i}o Negro, Argentina}
\affiliation{Instituto Balseiro, Universidad Nacional de Cuyo, Centro At\'omico Bariloche, 8400 San Carlos de Bariloche, R\'{\i}o Negro, Argentina}
\author{J. Milano}
\affiliation{CONICET, Centro At{\'{o}}mico Bariloche, 8400 San Carlos de Bariloche, R\'{\i}o Negro, Argentina}
\affiliation{Instituto Balseiro, Universidad Nacional de Cuyo, Centro At\'omico Bariloche, 8400 San Carlos de Bariloche, R\'{\i}o Negro, Argentina}
\author{S. A. Cannas}
\affiliation{Facultad de Matem\'atica, Astronom\'{\i}a y F\'{\i}sica, Universidad Nacional de C\'ordoba, Instituto de F\'{\i}sica Enrique Gaviola (IFEG-CONICET), Ciudad Universitaria, 5000 C\'ordoba, Argentina}

\begin{abstract}
In this work we studied a two dimensional ferromagnetic system using Monte Carlo simulations. Our model
includes exchange and dipolar interactions, a cubic anisotropy term, and uniaxial out-of-plane and in-plane ones.
According to the set of parameters chosen, the model including uniaxial out-of-plane
anisotropy has a ground-state which consists of a canted state with stripes of opposite
out-of-plane magnetization. When  the cubic anisotropy is introduced  zigzag patterns appear in the stripes
at fields close to the remanence.
An analysis of the anisotropy terms of the model shows that this configuration is related
to specific values of the ratio between the cubic and the effective uniaxial anisotropy.
The mechanism behind this effect is related to particular features of the anisotropy's
energy landscape, since  a global minima transition as a function of the applied field is
required in the anisotropy terms. This new mechanism for zigzags formation could be present in
monocrystal ferromagnetic thin films in a given range of thicknesses.

\end{abstract}

 \pacs{75.60.Ch,   
 	75.10.Hk,  
        75.70.Ak   
 	75.40.Mg}  
	
\date{\today}

\maketitle

\section{Introduction}

In ferromagnetic systems, modulated phases appear due to the competition between 
short-range exchange interactions and the unavoidable long-range dipolar ones.
In the particular case of thin films with strong out-of-plane anisotropy, this competition
produces a stripe phase at zero field; in this phase, parallel stripes with alternated
out-of-plane magnetization are formed. These kinds of patterns are usually found in magnetic
garnets \cite{Seul92PRA,Molho86JMMM,Molho87JAP}  and also in ultra-thin films, such as 
Fe on Cu \cite{PaKaHo1990,SaLiPo2010}.
Usually, in these systems a high out-of-plane field transforms the stripe phase in a
bubble phase. Under certain conditions magnetic garnets can also develop zigzags patterns
and other complex magnetics structures \cite{Seul92PRA,Molho86JMMM,Molho87JAP,Demand02JMMM}.

In the stripe phase, a magnetic field applied perpendicular to the film plane increases
the period of the stripes stretching  the thickness of the stripes aligned to the field
and shrinking the stripes pointing in the opposite direction \cite{Johansen13PRB}. In
some of these systems, a significant change in the stripe period is observed when either
the temperature or the magnetic field changes.
Using a smectic like model relying on effective local interactions energies such as bending
and compression, Sornette \cite{Sornette87JP}  proposed the following mechanism for
zig-zag formation.
In order to accommodate the  enhancement of the stripe period as the field is increased,
the system has to eject lines, i.e., domain walls, and this is conducted by the nucleation and
climbing of dislocations. However, when the field is decreased  and the stripe period shrinks,
the nucleation of a new  stripe by  edge dislocations is not observed. Instead, the system
develops an undulation  instability when a threshold in dilative strain is reached; a further
decrease of the field  transforms this sinusoidal undulation into a zigzag pattern.

In other systems with a reduced out-of-plane anisotropy, a canted phase can appear, i.e.,
in addition to the stripes with out-of-plane magnetization, an in-plane magnetization
component is  present \cite{Coisson08PRB,Coisson09JMMM,Sallica10PRB, Barturen12APL}.
In this canted spin configuration,
an in-plane  field parallel to the stripes should induce a stripe width
variation \cite{Saito64JPSJ}, however, this effect is difficult to be observed experimentally
and hitherto there are few experiments \cite{Lo70JAP,Talbi10JPCS}  showing the effect, aside from  certain cases in which an oscillating field is  needed in order to unpin the
stripes \cite{Lo70JAP}.

Recently, Barturen {\it et al.} \cite{Barturen12APL} have reported the presence of zigzags in
monocrystalline Fe$_{1-x}$Ga$_x$ thin films with a canted stripe configuration.
Since in these systems variations of the stripe width as function of the applied field are not
observed, the origin of the zigzag patterns should be based on a different
mechanism than that proposed by Sornette. In this work, we introduce and analyze
a simplified two-dimensional model which exhibits a canted stripe configuration \cite{Whitehead08PRB,Pighin12PRE}.
In this system, we study the magnetic pattern evolution under cycled in-plane applied fields.
We report a new mechanism for zigzag pattern formation which depends
on the ratio between the uniaxial out-of-plane anisotropy and the cubic anisotropy.
This mechanism does not assume stripe width variation; instead, it is based in the particular
form of the anisotropy energy landscape.

This paper is organized as follows: In Sec.~\ref{model}, we introduce the
Monte Carlo model and the numerical methods. In Sec.~\ref{results}, we
show the results of Monte Carlo simulations. In Sec.~\ref{anisotropy},
we analyze the anisotropy term of the energy in a single-spin approximation
to explain the Monte Carlo results. Finally, in Sec.~\ref{conclusions} we
summarize our results.

\section{Model}
\label{model}
\begin{figure}
\includegraphics[width=8cm]{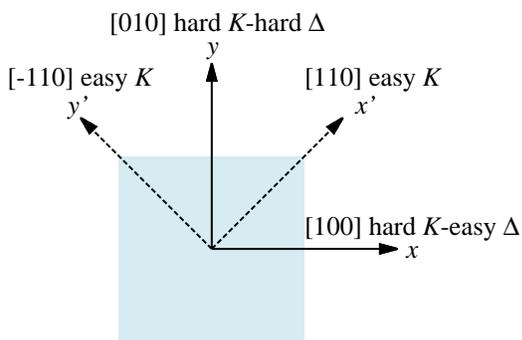}
\caption{Anisotropies easy axes scheme of Hamiltonian (\ref{eq1}).}
\label{fig:model}
\end{figure}

Our Monte Carlo simulations are ruled by the following two dimensional Heisenberg model:
\begin{widetext}
\begin{eqnarray}
\label{eq1}
 {\cal H} = - \mathcal{J} \sum_{\langle i,j \rangle} \vec{S}_i \cdot \vec{S}_j + \sum_{(i,j)} \left[ \frac{\vec{S}_i \cdot \vec{S}_j}{r^3_{ij}} - 3 \frac{(\vec{S}_i \cdot \vec{r}_{ij}) (\vec{S}_j \cdot \vec{r}_{ij})}{r^5_{ij}} \right] - \eta \sum_i (S^z_i)^2 \\ \nonumber
+ K \sum_i \left\{ \frac{1}{4} \left[(S^x_i)^2 - (S^y_i)^2 \right]^2 + (S^y_i)^2 (S^z_i)^2 +(S^z_i)^2 (S^x_i)^2 \right\}
+ \Delta \sum_i (S^y_i)^2 - \sum_i \vec{H} \cdot \vec{S}_i.
\end{eqnarray}
\end{widetext}
where $\vec{S}_i$ are dimensionless unit vectors, $\mathcal{J}$ is the exchange interaction strength,
$\eta$ is the out-of-plane anisotropy  constant, $K$ gives the strength of the cubic
magnetocrystalline anisotropy, and  $\Delta$ stands for an additionally two-fold in-plane anisotropy.
All the constants are  normalized relative to the dipolar coupling constant $\Omega$.\footnote{The dipolar constant
is $\Omega = \mu_0 (g\mu_B)^2$, where $g$ is the Lande factor, $\mu_0$ the vacuum
permitivity and $\mu_B$ the Bohr magneton}
$\langle i,j \rangle$ stands for a sum
over nearest-neighbors pairs of sites in a square lattice with $N=L_x \times L_y$
sites, $(i,j)$ stands for a sum over all pairs of sites, and $r_{ij}= |\vec{r}_i-\vec{r}_j|$
is the distance between sites $i$ and $j$.
In order to avoid lattice discretization effects
in the Monte Carlo  simulations,  the cubic anisotropy term
is rotated in $\pi/4$  with respect to an axis perpendicular to the plane.
In this way, the $[100]$ and $[010]$ are hard magnetization  directions (see Fig. \ref{fig:model}).
The additional  term corresponding to the factor $\Delta$  breaks the symmetry
of this two directions making $[010]$ harder as compared to the [100] direction.
This term is added because a breaking of the in plane four-fold cubic magnetic-crystalline
anisotropy has been observed in Fe$_{1-x}$Ga$_x$\cite{Barturen12APL} and
in Fe films \cite{Gustavsson02PRB} epitaxied over ZnSe buffers. This symmetry breaking
is associated to interfacial effects.

The numerical simulations were performed using a Metropolis algorithm with a single spin
update. The direction of each magnetic vector $\vec{S_i}$ is updated randomly in the unit sphere.
In all the simulations we start in a random spin configuration and then cool the  system with an
in-plane magnetic field pointing in a given crystallographic direction.
After that we cycle the field in the cooling direction to obtain the hysteresis loops.

The phase diagram of this model has been studied in the case of $\Delta=0$ and $K=0$ through Monte Carlo
simulations at finite temperature \cite{Carubelli08PRB, Whitehead08PRB, Pighin12PRE} and analytical 
calculation at zero temperature \cite{Po1998,YaGy1988,Pighin10JMMM}.
There is a region in the parameter space where the system shows a canted phase with
perpendicular striped patterns. This makes the model useful to study thin film systems with
a canted magnetic configuration.

In order to obtain a canted phase, we set the following 
parameters \cite{Pighin10JMMM, Pighin12PRE}: $\eta=7$, $\mathcal{J}=6$.
$K$ and $\Delta$ can be considered as small perturbations.
We choose $K=0.68$ and $\Delta=0.15$. These relatively small values ensure the system
remains in a canted state. We set $k_BT/\Omega = 0.2$ in all our analyses. This is a small
temperature since the ordering temperature is at least $40$ times larger.
The size of the system studied is $L_x=L_y=120$.

\section{Results}
\label{results}

\begin{figure}
\includegraphics[angle=-90,width=8cm,clip=true]{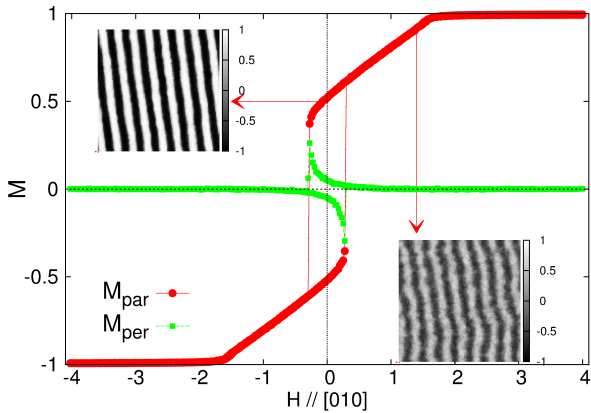}
\caption{(Color online) Hysteresis loops with $K=0.00$, $\Delta=0.00$, $k_BT/\Omega=0.2$, and the
field applied in the [010] direction. The snapshots show magnetic landscapes of the out
of plane  magnetization at remanence ($H=0$), and at $H=1.4$ close  to the appearance of
the stripes.}
\label{fig1:reference_loops}
\end{figure}

In Fig. \ref{fig1:reference_loops}, we show vectorial hysteresis \cite{Coisson08PRB} loops
simulated with $K=0$ and $\Delta = 0$ and the applied field in the in-plane [010]
direction. This corresponds to the  case where only the perpendicular uniaxial anisotropy
is present, and it is a useful reference for the analysis of the main results shown in the
following.
One can see the typical features observed in the hysteresis loops of materials with
perpendicular striped pattern with a canted magnetization, such as FePt \cite{Sallica10PRB} or
Fe$_{1-x}$Ga$_x$ \cite{Barturen12APL}.  At high saturating fields, the magnetization is in the plane
pointing in the direction of the applied field.  When the field is reduced, there is a characteristic
field at which stripes aligned to the field appear (see inset).
From this characteristic field at which the stripes appear down to the coercivity, the magnetization
inside the stripes continuously rotates. On one hand, as reflected in the vectorial hysteresis loop in Fig. \ref{fig1:reference_loops},
the in-plane rotation is marked by a linear behavior of the magnetization aligned to the field while the
perpendicular in-plane magnetization increases when the field is decreased reaching its maximum value
at coercivity. On the other hand, the out-of-plane magnetization goes up and down following the stripe pattern and increasing its absolute value, as can be observed through the increasing contrast of the stripes (see insets in Fig. \ref{fig1:reference_loops}).

\begin{figure}
\includegraphics[width=8cm,clip=true]{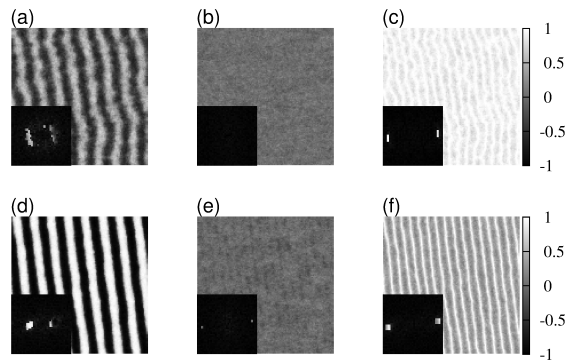}
\caption{Snapshots of magnetic patterns corresponding to two different applied
fields along the [010] in-plane direction.
Top panels correspond to $H=1.4$, and bottom panels to $H=0$.
(a) and (d) out-of-plane magnetization (see insets of Fig. \ref{fig1:reference_loops}),
(b) and  (e) in-plane magnetization perpendicular to the
applied field, and (c) and (f) magnetization parallel to the applied field.
Insets show the two-dimensional structure factor associated to each snapshot.}
\label{fig2:configurations}
\end{figure}

In Fig. \ref{fig2:configurations}, we show snapshots of the out-of-plane magnetization
patterns  already  shown in  the insets  of Fig. \ref{fig1:reference_loops} together
with patterns of the in-plane magnetization  parallel  and perpendicular to
the applied field. The upper panels correspond to $H=1.4$ and the  lower panels
to $H = 0$, i.e. the remnant state. The white lines of Figs. \ref{fig2:configurations}
(c) and (f) depict the domains walls. In our convention, white means positive and black negative, i.e., along and opposite to the field, respectively.
Since the in-plane magnetization perpendicular to the field --Figs. \ref{fig2:configurations} (b) and (e)-- does not show any
preferential orientation, the domain walls are of Bloch type.
The insets show the structure factor corresponding to each snapshot, defined as the squared modulus
of its Fourier transform. The two peaks observed on Figs. \ref{fig2:configurations} (a) and (d) account for the periodic structure and
are located at the characteristic wave-vector modulus $k^*=2 \pi/\lambda$, where $\lambda$ is the period
of the stripe pattern. In our case, $\lambda=15$ and $k^*=0.42$. The in-plane parallel magnetization shows
the presence of domain walls between two consecutive out-of-plane domains and thus has half the period of the stripe pattern.
Therefore, the characteristic wave vector in Figs. \ref{fig2:configurations} (c) and (f) is $2 k^*$, as shown in the insets.
Although difficult to observe under the present resolution, a weak perpendicular component can be detected in
the inset of Fig. \ref{fig2:configurations} (e), consistent with a Bloch domain wall (notice that the stripes are not completely
vertical).

\begin{figure}
\includegraphics[angle=-90,width=8cm,clip=true]{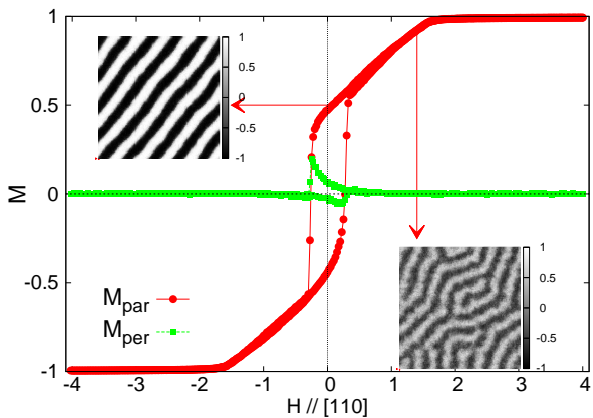}
\caption{(Color online) Hysteresis loops with $K=0.00$, $\Delta=0.00$, $k_BT/\Omega=0.2$, and the
field applied in the [110] direction. The snapshots show magnetic landscapes of the out
of plane  magnetization at remanence ($H=0$), and at $H=1.4$ close  to the apparition of the stripes.}
\label{fig1:reference_loops_b}
\end{figure}

When the field is rotated and applied in the $[110]$ direction, as shown in
Fig. \ref{fig1:reference_loops_b}, some differences can be observed
as compared to Fig. \ref{fig1:reference_loops} ($[010]$ direction).
In this case, the stripes start forming with several defects and
the hysteresis loop is slightly asymmetric; the descending branch being different from the ascending branch.
This asymmetry can be better visualized  through the hysteresis loop of the
perpendicular magnetization.
The slight difference between the loops in Figs. \ref{fig1:reference_loops} and \ref{fig1:reference_loops_b} arise from the spurious in-plane anisotropy introduced by lattice discretization effects in the numerical models used here. The underlying square lattice introduces a dependency of the domain-wall energy on the orientation of the stripes, which is particularly stressed in small systems.
Due to this effect, the [110] direction is magnetically slightly harder than the [010] and hence domain walls aligned along the lattice directions are favored. This mechanism is behind the observed defects on the stripe patterns in Fig. \ref{fig1:reference_loops_b} and is therefore responsible for the asymmetric loops.

\begin{figure}
\includegraphics[angle=-90,width=8cm,clip=true]{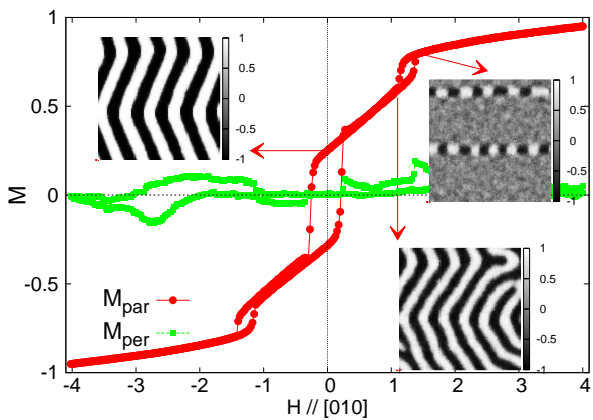}
\caption{(Color online) Hysteresis loops with $K=0.68$, $\Delta=0.00$, $k_BT/\Omega=0.2$, and the
field applied in the [010] direction.
The snapshots show perpendicular magnetic configurations corresponding to the upper branch of
the hysteresis loop: At the beginning of the minor loop
($H=1.4$), at the end ($H=1.1$), and at remanence ($H=0$).}
\label{fig3:loops_w017h}
\end{figure}

We turn now to the analysis of the effect of the cubic anisotropy. Since the cubic anisotropy term is rotated
in $\pi/4$, it counteracts the lattice effects we have observed in
Figs. \ref{fig1:reference_loops} and \ref{fig1:reference_loops_b}. In this
way, $[010]$ is a hard direction and $[110]$ is an easy direction and the lattice effect can be neglected.
Interestingly, as shown in the left inset of Fig. \ref{fig3:loops_w017h}, when a cubic anisotropy is
added to the model ($K=0.68$), zigzags in the stripe pattern appear at remanence.
In addition, at high fields, minor loops appear and we will show in the following that this is closely related to the zigzags
formation. When the field is decreased from saturation, two lines of bubbles of the out-of-plane magnetization form at a field which
correspond to the onset of the minor loops  ($H \sim 1.4$).
This magnetic configuration is shown in the upper right inset of Fig. \ref{fig3:loops_w017h}.
If the field is reduced to the end of the minor loops, these two lines of bubbles connect, forming
undulated stripes  with some defects (bottom right inset). Finally, at remanence the undulated stripes take the form
of  well-defined  zigzags.
Note that the stripe width slightly changes as the field is decreased, being it larger
at remanence. This is not related to the zigzag apparition, as in the case of the mechanism
proposed by Sornette, because once the stripes appear they are already undulated.

\begin{figure}
\includegraphics[width=8cm,clip=true]{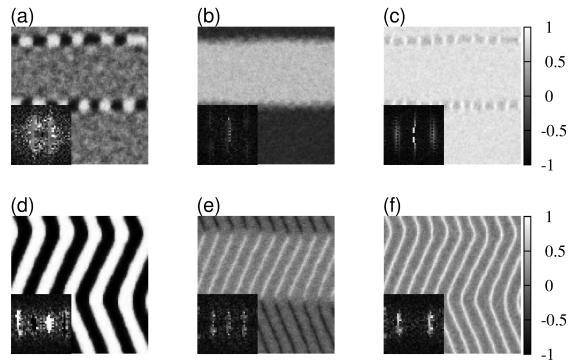}
\caption{Snapshots of magnetic patterns corresponding to two different applied
fields along the [010] in-plane direction.
Top panels correspond to $H=1.4$, and bottom panels to $H=0$.
(a) and (d) out-of-plane magnetization (see two top insets of Fig. \ref{fig3:loops_w017h}),
(b) and  (e) in-plane magnetization perpendicular to the
applied field, and (c) and (f) magnetization parallel to the applied field.
The two-dimensional structure factor associated to each snapshot are presented in the insets, which show signatures of the different periodic structures.}
\label{fig3b:configurations}
\end{figure}

Typical magnetic configurations associated to this process can be seen in
more detail in Fig. \ref{fig3b:configurations}. The magnetization inside the bubbles alternates in the out-of-plane
direction and is canted in  the direction of applied field, Figs. \ref{fig3b:configurations} (a) and (c),
respectively. At the interface between bubbles of different orientations, the in-plane magnetization
points in the direction of the applied field, small white threads in Figs. \ref{fig3b:configurations} (c).
The in-plane magnetization perpendicular to the applied field arranges into
domains (two in this case due to the size of the system) which point in opposite directions as indicated by the
dark and light gray regions in Fig. \ref{fig3b:configurations} (b). The interface between perpendicular magnetization domains
is mediated by bubble lines which can be considered as wide domain walls with a complex internal structure.
The presence of these domains reduces the dipolar energy of the in plane magnetization component.
Since dipolar interactions are  minimized by in-plane configurations,
the energy increment due to the creation of the bubble lines
should  be small in order to compensate the dipolar energy reduction.
It is known that Bloch's domain walls are favored in two-dimensional systems \cite{Po1998}.
Because of this fact, when the field is decreased and the stripes emerge, they follow the
orientation of the in-plane magnetization.
In other words, the orientation of the stripes depends on the orientation of the in-plane
magnetization of the domains at which they arise.
According to this, the corners of the zigzags correspond to the bubble lines, i.e., these are the
lines at which  stripes of different orientations connect.
At zero field, Figs. \ref{fig3b:configurations} (d), (e) and (f), the domains of
perpendicular in-plane magnetization disappear. Now, the in-plane magnetization inside
the domain walls follows the orientation of the stripes. This is evidenced in
Fig. \ref{fig3b:configurations} (e) where the in-plane component
of the magnetization inside the domain walls has a different sign depending
on the orientation of the stripes.

\begin{figure}
\includegraphics[angle=-90,width=8cm,clip=true]{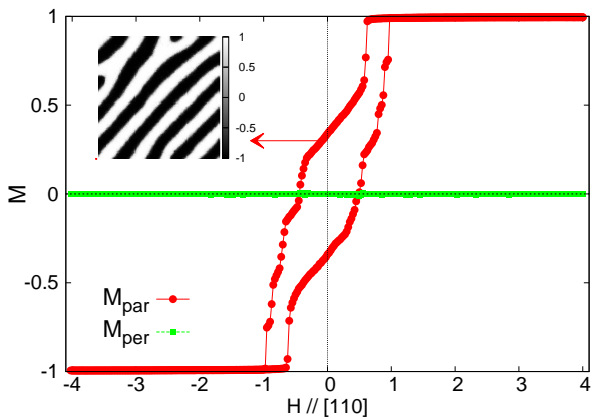}
\caption{(Color online) Hysteresis loops obtained with $K=0.68$, $\Delta=0.00$, $k_BT/\Omega=0.2$. The field is applied in the $[110]$
direction. The snapshot shows  perpendicular magnetic configurations corresponding to the upper branch of the hysteresis
loop  at remanence ($H=0$).}
\label{fig4:loops_w017e}
\end{figure}

Figure \ref{fig4:loops_w017e} shows hysteresis loops with the same parameters used in Fig. \ref{fig3:loops_w017h},
but now the field is applied in the $[110]$ direction, i.e., the easy-$K$ direction. We see that the mechanism operating  in
the magnetization process from saturation to remanence is different as compared to that of the
$[010]$ hard-$K$ direction. The reversible part of the loop observed after the appearance
of the stripes in Figs. \ref{fig1:reference_loops} and \ref{fig3:loops_w017h}  is not
present in this case. On the other hand, the in-plane magnetization perpendicular to
the field is always zero, indicating that the magnetization goes to the out-of-plane direction
before the inversion and does not rotate in the plane. In this applied field direction,
zigzags patterns  are not observed; instead, some defects like dislocations can be obtained, as the one
shown at remanence (see inset of Fig. \ref{fig4:loops_w017e}).

\begin{figure}
\includegraphics[angle=-90,width=8cm,clip=true]{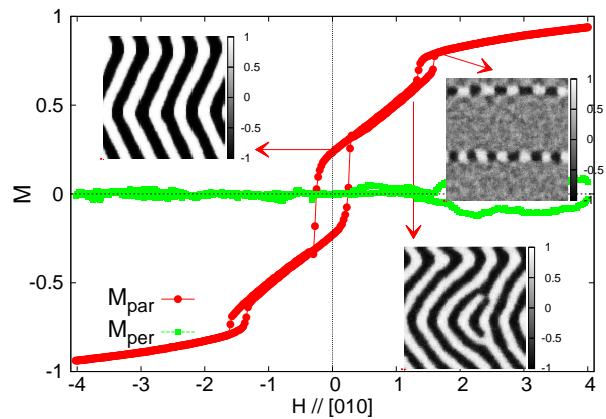}
\caption{(Color online) Hysteresis loops obtained with $K=0.68$, $\Delta=0.15$, $k_BT/\Omega=0.2$, the field is applied in the [010]
direction. The snapshots show  perpendicular magnetic configurations corresponding to the upper branch of the hysteresis
loop. At the beginning of the minor loop ($H=1.6$), at the end ($H=1.27$) and at remanence ($H=0$).}
\label{fig5:loops_w017D015}
\end{figure}

At this point, one might think that the symmetry between the hard-$K$ directions ($[100]$ and $[010]$) is one of the keys in the formation of the zigzag pattern. We therefore investigated whether the zigzag formation can be unfavored by a small perturbation making $[100]$ and $[010]$ directions energetically different.
As shown in Fig. \ref{fig5:loops_w017D015}, similar hysteresis loops to those shown in Figs. \ref{fig3:loops_w017h} and \ref{fig4:loops_w017e}
are found when the in-plane uniaxial anisotropy is present ($\Delta=0.15$),
breaking the four-fold in-plane anisotropy of the cubic term.
Now, the $[010]$ direction, along which the field is applied, is harder than the  $[100]$ direction due to the presence of the $\Delta$ term; the easy-$K$ directions $[110]$ and $[-110]$ continue being equivalent (see Fig. \ref{fig:model}). The minor loops shift toward
higher fields  but the  phenomenology is quite similar to the one in Fig. \ref{fig3:loops_w017h}, as observed in the insets.
If the field is applied in $[100]$ direction, (not show here), the zigzags are still observed
at remanence but its period changes. This change is related
to the difference in the energy of the bubbles lines induced by the presence of uniaxial
in-plane term. Finally, in this case, as in Fig. \ref{fig4:loops_w017e}, when the field is applied in
the $[110]$, the zigzags do not form.

\section{Anisotropy analysis}
\label{anisotropy}

In this section, we analyze the anisotropy term of the Hamiltonian, (\ref{eq1}),
in a single spin approximation. For simplicity, we refer the cosine directors to the in-plane rotated frame, indicated by $x'$ and $y'$ in Fig. \ref{fig:model}. Since we want to study the appearance of the zigzag pattern, we particularly focus on the case where the external field $H$ is oriented in the $[010]$ direction, which corresponds here to the diagonal of the $x'-y'$ coordinate system and thus implies an equal contribution from $\alpha_1$ and $\alpha_2$. Therefore, the single-spin anisotropy energy can be expressed as
\begin{eqnarray}
\label{eq2}
E &=& K(\alpha_1^2\alpha_2^2 + \alpha_1^2\alpha_3^2+\alpha_2^2\alpha_3^2) + \frac{\Delta}{2}(\alpha_1+\alpha_2)^2 \nonumber \\
& &- \frac{H}{\sqrt{2}}(\alpha_1+\alpha_2) -\eta_e \alpha_3^2,
\end{eqnarray}
where $\alpha_i$ are cosine directors with respect to the in-plane easy-$K$ directions (see Fig. \ref{fig:model}) and satisfy $\alpha_1^2 + \alpha_2^2 + \alpha_3^2=1$.
The effective  uniaxial anisotropy $\eta_e$ takes into account the dipolar
(shape anisotropy) and the uniaxial anisotropy ($\eta$) in the Hamiltonian of Eq. (\ref{eq1}).
We want to analyze the evolution with the external field $H$ of the  absolute energy minima which at $H=0$ are located at
$\alpha_3^0=\pm1$ and $\alpha_1^0 =\alpha_2^0=0$, i.e., with the magnetization fully oriented out of plane
\footnote{In the lamellar phase with
high out-of-plane anisotropy, the spins inside the stripes are accommodated in these two effective minima}.

\subsection{Analysis of the critical points}

From  Eq. (\ref{eq2}) and using that $1- \alpha_1^2 - \alpha_2^2 = \alpha_3^2$, we obtain an expression for the energy that only depends on $\alpha_1$ and $\alpha_2$:
\begin{eqnarray}
\label{eq3}
E &=& K(\alpha_1^2 + \alpha_2^2 - \alpha_1^2\alpha_2^2 - \alpha_1^4 - \alpha_2^4) + \frac{\Delta}{2}(\alpha_1+\alpha_2)^2 \nonumber \\
& & - \frac{H}{\sqrt{2}}(\alpha_1+\alpha_2) -\eta_e (1 - \alpha_1^2 - \alpha_2^2),
\end{eqnarray}
provided  that $\alpha_1^2 + \alpha_2^2 \le 1$. A graphical inspection of this energy model shows that all the minima (for the present range of parameters values) satisfy either of the following conditions: (a) $\alpha_1^0=\alpha_2^0$; (b) $\alpha_3^0=0$. The values $\alpha_1^0$ and $\alpha_2^0$ which minimize the energy model, Eq. (\ref{eq3}), are obtained through its partial derivatives, given by
\begin{eqnarray}
\frac{\partial E}{\partial \alpha_1} & = & [2K(1-\alpha_2^2) + 2 \eta_e ] \alpha_1 + \Delta (\alpha_1+\alpha_2) \nonumber \\
& & - \frac{H}{\sqrt{2}} - 4 K \alpha_1^3 = 0, \label{eq4} \\
\frac{\partial E}{\partial \alpha_2} & = & [2K(1-\alpha_1^2) + 2 \eta_e ] \alpha_2 + \Delta (\alpha_1+\alpha_2) \nonumber \\
& & - \frac{H}{\sqrt{2}} - 4 K \alpha_2^3 = 0. \label{eq4b}
\end{eqnarray}
In order to study the stability of the solutions of this set of equations, the second derivatives of Eq. (\ref{eq3}) should also be considered:
\begin{eqnarray}
\label{eq5}
\frac{\partial^2 E}{\partial \alpha_1 \partial \alpha_2} = -4K\alpha_1 \alpha_2 +  \Delta, \\
\frac{\partial^2 E}{\partial \alpha_1^2} = 2K(1-\alpha_2^2) + 2 \eta_e +  \Delta - 12 K \alpha_1^2, \\
\frac{\partial^2 E}{\partial \alpha_2^2} = 2K(1-\alpha_1^2) + 2 \eta_e +  \Delta - 12 K \alpha_2^2.
\end{eqnarray}
In the following we analyze the solutions of Eqs.~(\ref{eq4}) and (\ref{eg5}) in order to obtain the different critical points describing the magnetization evolution observed, for example, in Fig.~\ref{fig3:loops_w017h}.

\subsubsection{Symmetric case: $\alpha_1=\alpha_2=\alpha$}

\begin{figure}
\includegraphics[angle=0,width=7cm,clip=true]{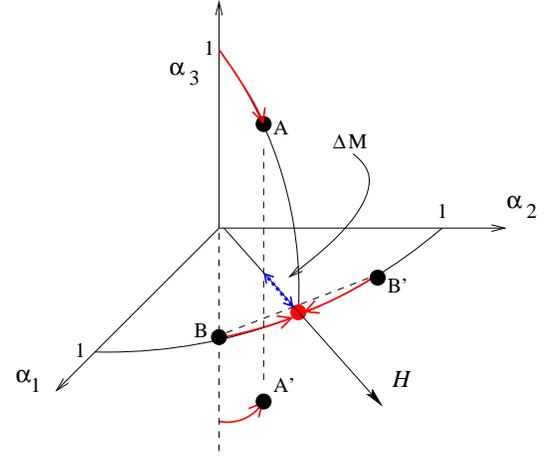}
\caption{(Color online) Scheme of evolution of the minima with an increasing external field (red arrows). The four minima, $A$, $A'$ and $B$, $B'$, at the transition at $h_1^*$ are indicated with black dots. The dashed double arrow indicates the increase $\Delta M$ of the magnetization associated to the transition at $h_1^*$. At $h_2^*$ the two minima with $\alpha_1^0 \neq \alpha_2^0$ collapse onto the $\alpha_1^0 = \alpha_2^0=1/\sqrt{2}$ red point.}
\label{fig:minima}
\end{figure}

Since when $H=0$ the energy has two absolute minima at $\alpha_1^0=\alpha_2^0=0$ and $\alpha_3^0=\pm 1$, i.e., with the magnetization
perpendicular to the film plane, we expect that for small applied fields ($[010]$ direction), these minima will move
in the field direction. When $\alpha_1=\alpha_2 = \alpha$,  Eqs. (\ref{eq4}) and (\ref{eq4b}) reduce to the following condition:
\begin{equation}
\label{eq6}
P(\alpha) = -3 \alpha^3 + (1 + \xi  +  \delta)\alpha = \frac{h}{\sqrt{8}},
\end{equation}
where $\delta = \frac{\Delta}{K}$, $\xi=\frac{\eta_e}{K}$, and $h=\frac{H}{K}$. The solutions of the above equation are given by the intersection between the cubic polynomial $P(\alpha)$ and the horizontal
line corresponding to the applied field. At $h=0$, the only stable minimum of the energy is the $\alpha^0=0$ solution. When $h$ increases, the value of $\alpha^0(H)$ corresponding to this minimum goes to positive values. \footnote{The other two solutions of the cubic equation correspond to maxima.} Since $P(\alpha)$ has a local maximum at positive values of $\alpha$, the value of $\alpha$ at this maximum is the upper limit that $\alpha^0(H)$ can take
as the field increases. This value is
\begin{eqnarray}
\label{eq7}
\alpha_{max} =  \frac{\sqrt{1 + \xi + \delta}}{3}.
\end{eqnarray}
The value of the critical field necessary to be applied so the minimum of the energy is at $\alpha_{max}$ is
\begin{eqnarray}
\label{eq8}
h_1^* = \frac{H_1^*}{K} = \sqrt{8} P(\alpha_{max}) = \frac{4 \sqrt{2}}{9}(1 + \xi + \delta)^{3/2}.
\end{eqnarray}
Therefore, $\alpha^0(H_1^*)=\alpha_{max}$ and for fields in the range $0 < H < H_1^*$, the energy is minimized at $0 < \alpha^0 < \alpha^0(H_1^*)$ and $\alpha_3^0=\pm \sqrt{1-2 (\alpha^0)^2}$. These solutions are represented schematically in Fig.~\ref{fig:minima} as the $A$ and $A$ points.

Let us analyze the stability of these minima as they move toward  the direction of the
applied field. The second derivatives of Eq. (\ref{eq3}) with $\alpha_1=\alpha_2=\alpha$ are
\begin{eqnarray}
\label{eq9}
\frac{\partial^2 E}{\partial \alpha_1 \partial \alpha_2} = -4K\alpha^2 +  \Delta, \\
\frac{\partial^2 E}{\partial \alpha_1^2} = 2(K + \eta) + \Delta -14K\alpha^2.
\end{eqnarray}
Using, these expressions we can obtain the Hessian. At the point where the Hessian
is zero,  the minimum or maximum becomes a saddle point. Two solutions are obtained:
\begin{eqnarray}
\label{eq10}
\alpha_+ = \frac{\sqrt{1+\xi}}{\sqrt{5}}, \\
\label{eq11}
\alpha_- = \frac{\sqrt{1+\xi + \delta}}{3}.
\end{eqnarray}
Note that $\alpha_- = \alpha_{max} < \alpha_+$, provided that $\delta$ is small.
Then, once $\alpha^0$ reach the value of $\alpha_{max}$  the minima become unstable.

\subsubsection{In-plane case: $\alpha_3=0$}

As the field is increased, two other local minima with $\alpha_1^0 \neq \alpha_2^0$ and $\alpha_3^0=0$ appear, namely,
in-plane solutions not aligned with the field. These solutions are located symmetrically with respect to the direction of the
field (see Fig.\ref{fig:minima})  and we call them $B$ and $B'$. When the solutions $A$ and $A'$ loose stability the solutions
$B$ and $B'$ become the absolute minima. As the field further increases, the solutions $B$ and $B'$ converges to a single
one aligned with the field ($\alpha_1^0 = \alpha_2^0 =1/\sqrt{2}$ and $\alpha_3^0=0$).

We shall now find the critical field $h_2^*$ at which the two in-plane minima join.
Taking $\alpha_3=0$ and $\alpha'=\alpha_1=\sqrt{1-\alpha_2^2}$,  Eqs. (\ref{eq4}) and (\ref{eq4b}) reduce to
\begin{eqnarray}
 \label{eq12}
 Q(\alpha') = & & \alpha' (1-2 \alpha'^2) \nonumber \\
 &+& \frac{h \alpha' +  \sqrt{2} \delta (1- 2 \alpha'^2)}{\sqrt{8(1-\alpha'^2)}} = \frac{h}{\sqrt{8}}.
\end{eqnarray}
This equation has three real solutions. One corresponds to $\alpha_0'^0=1/\sqrt{2}$, i.e., $\alpha_1^0 = \alpha_2^0 = 1/\sqrt{2}$.
The other two symmetric solutions are $(\alpha_1^0,\alpha_2^0) = (\alpha_+'^0,\alpha_-'^0)$ and $(\alpha_1^0,\alpha_2^0) = (\alpha_-'^0,\alpha_+'^0)$, with $\alpha_+'^0 < 1/\sqrt{2}$ and $\alpha_-'^0 = \sqrt{1-{\alpha_+'^0}^2}$.
For $h<h_2^*$, the solution $\alpha_0'^0$ is a maximum and the other two solutions are minima.
When increasing the field, these two minima converge to $\alpha_0'^0$ which becomes the stable solution to Eq.~(\ref{eq12}),
meaning that the magnetization  is fully aligned with the external field and saturated in plane (see Fig.~\ref{fig:minima}).
Using the bordered Hessian matrices for the constrained extrema problem, we analyze the stability 
of these minima (see Appendix), and we obtain the critical field at which the two in-plane minima join:
\begin{equation}
\label{eq13}
h_2^* = \frac{H_2^*}{K} = 2 + 2 \delta .
\end{equation}
If $h_1^* < h < h_2^*$, the two in-plane symmetric solutions exist and this is a condition for the existence
of the zigzag pattern.
From this condition, we obtain
\begin{equation}
\label{eq14}
\xi <  \frac{3}{2}3^{1/3} ( 1 + \delta )^{2/3} - ( 1 + \delta ).
\end{equation}
This gives a relation between anisotropy constants $K$, $\eta_e$, and $\Delta$ for 
the existence of the zigzag pattern. In particular, for $\Delta =0$, one has that $\eta_e < 1.16 K$.

\section{Summary and final remarks}
\label{conclusions}

In the following, we shall describe the whole scenario that emerges from the previous model 
(Eq.~(\ref{eq2})), and for simplicity we will focus on the case $\Delta =0$. 
Figure~\ref{fig:minima} shows a scheme of the energy minima in the 
$\alpha_1,\alpha_2,\alpha_3$ space. When $H=0$ the magnetization is fully out of plane: 
$\alpha_3^0=\pm 1$ and $\alpha_1^0=\alpha_2^0=0$. When increasing the field in
the [010] direction, and for $0<H<H_1^*= K (1+\eta_e/K)^{3/2} 4 \sqrt{2}/9$, the 
magnetization still has an out-of-plane component and is canted in the direction of 
the field, with $|\alpha_3^0| > 0$ and $\alpha_1^0=\alpha_2^0>0$. At the field $H_1^*$ the
magnetization along the external field, $\alpha_1^0=\alpha_2^0$, is no longer stable and 
now the magnetization has two in-plane symmetric states given by $\alpha_3^0=0$ and 
$(\alpha_1^0,\alpha_2^0) = (\alpha_+'^0,\alpha_-'^0)$ and 
$(\alpha_1^0,\alpha_2^0) = (\alpha_-'^0,\alpha_+'^0)$. By further increasing the external 
field, the projection of these two magnetization states along the field increases until the 
magnetization finally aligns with the field at the value $H=H_2^*= 2 K$.
For $H > H_2^*$, the magnetization is saturated in the direction of the external field.

\begin{figure}
\includegraphics[angle=-90,width=7cm,clip=true]{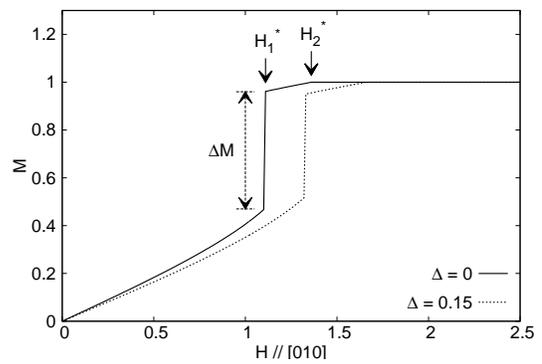}
\caption{Evolution of the projection of the magnetization with the external field applied in the $[010]$ direction. Two curves, corresponding to $\Delta=0$ and $0.15$, are shown. The characteristic fields $H_1^*$ and $H_2^*$, as well as the increase of the in-plane magnetization $\Delta M$ at $H_1^*$, are shown.}
\label{fig:mindirH}
\end{figure}

When the external field is in the range $H_1^* < H < H_2^*$ there are, in the single-spin 
approximation, two equivalent in-plane magnetization states, not aligned with the external 
field. These two states can be observed in Figs.~\ref{fig3b:configurations}(b) and (c). 
The line of bubbles observed in Fig.~\ref{fig3b:configurations}(a) are the domain walls 
between the two in-plane magnetization states. The bubble structure of these domain walls 
is the result of the dipolar energy term, not present in the single-spin approximation, 
and are responsible for the origin of the zigzag pattern. At smaller external field values, 
$H<H_1^*$, canted magnetization states with an out-of-plane component, such as the ones in the 
bubbles, are favored. These are the states inside the domains observed in 
Figs.~\ref{fig3b:configurations}(a)--(c). The zigzag pattern then results from the connection 
of the bubbles when the canted states are preferred.
In Fig.~\ref{fig:mindirH}, we plot the component of the magnetization in the direction of 
the applied field $H$, i.e  $M=(\alpha_1+\alpha_2)/\sqrt{2}$. The two critical fields 
$H_1^*$ and  $H_2^*$ are shown.

According to the previous scenario, the fields $H_1^*$ and $H_2^*$ can be identified in the single spin model with 
the characteristic field values of the minor loop and the saturation field, respectively. 
In order to compare the predictions of the single spin approximation with the Monte Carlo 
simulations we choose the following set of parameters for Eq. (\ref{eq2}): $\eta_e=0.72$ 
and $K=0.68$.
With these parameters we emulate the anisotropy terms of the model Hamiltonian 
(Eq. (\ref{eq1})). As an approximation, the effective uniaxial anisotropy $\eta_e$ 
introduced in order to take into account the dipolar shape anisotropy is computed as the sum
of uniaxial anisotropy $\eta$ and the effective planar dipolar anisotropy.
The effective planar anisotropy correspond to the value of the anisotropy at which the system
undergoes a planar to perpendicular reorientation (see Ref. \onlinecite{Pighin10JMMM}).
Note that the whole set of parameters satisfy Eq. (\ref{eq14}).
The  minor loops observed in Figs.~\ref{fig3:loops_w017h} and \ref{fig5:loops_w017D015}
are related to $H_1^*$ in the single spin model.
The values of $H_1^*$ obtained as the average value between borders of the 
minor loops of Figs.~\ref{fig3:loops_w017h} and \ref{fig5:loops_w017D015} are 
in a very good agreement with the values obtained in the single spin model. 
For the cases $\Delta = 0$ and $0.15$, the values $H_1^* \sim 1.25$ (Fig.~\ref{fig3:loops_w017h}) 
and $H_1^* \sim 1.47$ (Fig.~\ref{fig5:loops_w017D015}) were obtained with numerical 
simulations, while the single spin model predictions for each case are $H_1^* = 1.26$ and $1.47$.
We see that the values predicted by the single-spin approximation for $H_1^*$ agree very 
well with the values coming from Monte Carlo simulations. This indicates that the chosen 
value for $\eta_e$ accurately describes the numerical data and that the single-spin 
approximation gives a good description of the transition occurring at $H_1^*$ and the 
appearance of the zigzag patterns. 
However, the values predicted for the saturation field  $H_2^*$ do not agree 
with the  numerical simulations.  
This points to the limitations of the single-spin model to take into account
thermal fluctuation and also to the fact that dipolar interactions are not accurately
described by an effective anisotropy when the magnetization is mainly in the plane.

Summarizing, the zigzag mechanism that emerges from the present analysis is a direct 
consequence of cubic anisotropy, which gives rise to two pairs of effective local minima 
that exchange stability as the field changes. For instance, the appearance of a bubble
state depends on the ratio between the cubic anisotropy and the effective uniaxial 
anisotropy that takes into account dipolar energy contributions. The single absolute 
minimum at high fields transforms, as the field is decreased; first, in two minima with 
the magnetization in the film plane, and then, by a further reduction of the field, in 
two minima with out-of-plane magnetization. Close to the transition from two absolute 
minima in the plane to two absolute minima out of plane, the energies of these four minima 
are similar. The  proximity between the energy of the minima allows the formation of bubble
lines without paying so much energy, which in turn produces a reduction of the dipolar
energy and the appearance of the zigzag patterns.

\acknowledgements
S.B. acknowledges partial support by CONICET Grant No. PIP11220090100051. J.M. and M. B. acknowledge 
partial support by CONICET through PIP grant No. 112200901 00258 and ANPCyT through PICT grant No. 2010-0773. 
O.V.B and S.A.C.  acknowledge partial support by CONICET through PIP grant No. 11220110100213 and grants from 
SeCyT, Universidad Nacional de C\'ordoba (Argentina).

\appendix*
\section{Constrained critical point analysis}

In this appendix we provide details of the calculations on the stability analysis of the critical
points in the constrained energy problem.

Let us consider the energy $E$ and the constraint $g(\alpha_1, \alpha_2, \alpha_3)=0$,

\begin{eqnarray}
\label{eq:H}
E &=& K(\alpha_1^2\alpha_2^2 + \alpha_1^2\alpha_3^2+\alpha_2^2\alpha_3^2) \nonumber
 + \frac{\Delta}{2}(\alpha_1+\alpha_2)^2 \\  
  & &- \frac{H}{\sqrt{2}}(\alpha_1+\alpha_2) -\eta_e \alpha_3^2 \\
 g &=& 1 - \alpha_1^2 - \alpha_2^2 - \alpha_3^2 = 0.
\end{eqnarray}
Then the Lagrangian function $L$ of the problem is:
\begin{equation}
\label{lagrange}
L = E - \lambda g
\end{equation}
where $\lambda$ is the Lagrange multiplier.
The critical points $\alpha_1^*$, $\alpha_2^*$, $\alpha_3^*$ and $\lambda^*$ of the Lagrangian function
are solutions of the following equations:

\begin{eqnarray}
\label{critical}
\frac{\partial L}{\partial \alpha_1} &=& 2K \alpha_1(\alpha_2^2 + \alpha_3^2) 
 + \Delta(\alpha_1+\alpha_2) \\ \nonumber
                                      & &- \frac{H}{\sqrt{2}} + 2 \lambda \alpha_1 = 0\\
\frac{\partial L}{\partial \alpha_2} &=& 2K \alpha_2(\alpha_1^2 + \alpha_3^2) 
 + \Delta(\alpha_1+\alpha_2) \\ \nonumber
                                      & &- \frac{H}{\sqrt{2}} + 2 \lambda \alpha_2 = 0 \\
\frac{\partial L}{\partial \alpha_3} &=& 2K \alpha_3(\alpha_1^2 + \alpha_2^2)
 - 2 \eta_e \alpha_3  + 2 \lambda \alpha_3 = 0 \\
 \frac{\partial L}{\partial \lambda} &=&(\alpha_1^2 + \alpha_2^2 + \alpha_3^2)-1=0 \label{vinculo}
\end{eqnarray}
In order to classify the critical points, we have to analyze the determinants of the {\it bordered Hessian matrices} $H_4$ and $H_3$
evaluated  at the  critical points ($\alpha_1^*$, $\alpha_2^*$, $\alpha_3^*$, and $\lambda^*$). These matrices are:

\begin{equation}
H_4=\left( \begin{array}{cccc}
0 & -g_x & -g_y & -g_z\\
-g_x & L_{xx} &  L_{xy} &  L_{xz} \\
-g_y &  L_{yx} &  L_{yy} &  L_{yz}\\
-g_z &  L_{zx} &  L_{zy} &  L_{zz} \end{array} \right),   
\end{equation}
and if $ g_x(\alpha_1^*,\alpha_2^*,\alpha_3^*) \ne 0 $ and/or  $g_y(\alpha_1^*,\alpha_2^*,\alpha_3^*) \ne 0$ 
\begin{equation}
H_3=\left( \begin{array}{ccc}
0 & -g_x & -g_y \\
-g_x & L_{xx} &  L_{xy}  \\
-g_y &  L_{yx} &  L_{yy} \end{array} \right).  
\end{equation}
Here, $g_x=\frac{\partial g}{\partial \alpha_1}$ , $g_y=\frac{\partial g}{\partial \alpha_2}$ and $g_z=\frac{\partial g}{\partial \alpha_3}$.
Similarly, the double subscript in $L$ refers to the second partial derivatives, for instance,
$L_{xy}=\frac{\partial ^2 L}{\partial \alpha_1 \partial \alpha_2 }$

\begin{itemize}
  \item if $-det(H_4)>0$ and  $-det(H_3)>0$  the critical point is a minimum.
  \item if $-det(H_4)>0$ and  $-det(H_3)<0$  the critical point is a maximum.
  \item if $-det(H_4)<0$ then critical point is a saddle point.
\end{itemize}

In our problem the Hessian matrices are:
\begin{widetext}
\begin{equation}
H_4=\left( \begin{array}{cccc}
0 & 2\alpha_1 & 2\alpha_2 & 2\alpha_3 \\
2\alpha_1 & 2K(\alpha_2^2 + \alpha_3^2) + \Delta + 2 \lambda &  4K\alpha_1\alpha_2 + \Delta& 4K\alpha_1\alpha_3 \\
2\alpha_2 &  4K\alpha_1\alpha_2 + \Delta &  2K(\alpha_1^2 + \alpha_3^2) + \Delta + 2 \lambda  & 4K\alpha_2\alpha_3\\
2\alpha_3 &  4K\alpha_1\alpha_3 &  4K\alpha_2\alpha_3 &  2K(\alpha_1^2 + \alpha_2^2)+2(\lambda-\eta_e) \end{array} \right)   
\end{equation}
and
\begin{equation}
H_3=\left( \begin{array}{cccc}
0 & 2\alpha_1 & 2\alpha_2 \\
2\alpha_1 & 2K(\alpha_2^2 + \alpha_3^2) + \Delta + 2 \lambda &  4K\alpha_1\alpha_2 + \Delta \\
2\alpha_2 &  4K\alpha_1\alpha_2 + \Delta &  2K(\alpha_1^2 + \alpha_3^2) + \Delta + 2 \lambda \end{array} \right)   
\end{equation}
\end{widetext}
We would like to do the stability analysis for the critical point with $\alpha_1^* = \alpha_2^* = \frac{1}{\sqrt{2}}$, $\alpha_3^*=0$ 
and $\lambda = \lambda^*$. In order for this to be a critical point, and using Eq.~(\ref{critical}) one obtains that the 
Lagrange multiplier is field dependent,
\begin{equation}
 \lambda^* = \frac{H}{2} - \frac{K}{2} - \Delta.
\end{equation}
Then evaluating the determinants of $H_4$ and $H_3$ for this critical point we obtain
\begin{equation}
-det(H_4) = -det(H_3)[H + K - 2(\Delta + \eta_e)]
\end{equation}
and
\begin{equation}
-det(H_3) = 4[H - 2(K + \Delta)]
\end{equation}
Equating to zero the first and second factors in $-det(H_4)$ 
we get:
\begin{eqnarray}
 H_2^* = 2 (K + \Delta) \\
 H_3^* = 2 (\Delta + \eta_e -K/2)
\end{eqnarray}

If $\eta_e < \eta_e^* = 3K/2$ then we are in the {\em strong dipolar regime} where $H_3^* < H_2^*$. Then,
\begin{itemize}
 \item if $H < H_3^*$ we have that $-det(H_4) > 0$ and $-det(H_3) < 0$ and thus the critical point is a maximum,
 \item if $H_3^* < H < H_2^*$ then $-det(H_4) < 0$ and the critical point is a saddle point,
 \item if $H_2^* < H$ we have that $-det(H_4) > 0$ and $-det(H_3) > 0$ and thus the critical point is a minimum.
\end{itemize}
Since $\eta_e \approx K < \eta_e^*$, this is the case we are interested in and we have a well defined $H_2^*$ field value.

On the other hand, if $\eta_e > \eta_e^* = 3K/2$ then we are in the {\em weak dipolar regime} where $H_3^* > H_2^*$. Then:
\begin{itemize}
 \item if $H < H_2^*$ we have that $-det(H_4) > 0$ and $-det(H_3) < 0$ and thus the critical point is a maximum,
 \item if $H_2^* < H < H_3^*$ then $-det(H_4) < 0$ and the critical point is a saddle point,
 \item if $H_3^* < H$ we have that $-det(H_4) > 0$ and $-det(H_3) > 0$ and thus the critical point is a minimum.
\end{itemize}
Then, in this case we have to go beyond the field $H_3^*$ (which is larger than $H_2^*$) in order to have a minimum critical point. 
In this weak dipolar regime the external in-plane field has to win over the weak dipolar contribution in order to generate a fully 
in-plane magnetic moment.


\end{document}